\documentclass[twocolumn,aps,prx,bm,amsmath,superscriptaddress]{revtex4-1}
\usepackage{amssymb}
\usepackage{mathrsfs}
\usepackage{graphicx,graphics,subfigure}
\usepackage{epstopdf}
\usepackage{epsfig}
\usepackage{float}
\usepackage{txfonts}
\usepackage{xcolor}

\usepackage{braket}

\begin{document}
\title{Observation of a dynamical quantum phase transition by a superconducting qubit simulation}

\author{Xue-Yi Guo}

\affiliation{Institute of
Physics, Chinese Academy of Sciences, Beijing 100190, China}
\affiliation{School of Physical Sciences, UCAS, Beijing 100190, China}

\author{Chao Yang}

\affiliation{Institute of
Physics, Chinese Academy of Sciences, Beijing 100190, China}
\affiliation{School of Physical Sciences, UCAS, Beijing 100190, China}

\author{Yu Zeng}

\affiliation{Institute of
Physics, Chinese Academy of Sciences, Beijing 100190, China}
\affiliation{School of Physical Sciences, UCAS, Beijing 100190, China}

\author{Yi Peng}

\affiliation{Institute of
Physics, Chinese Academy of Sciences, Beijing 100190, China}
\affiliation{School of Physical Sciences, UCAS, Beijing 100190, China}

\author{He-Kang Li}
\affiliation{Institute of
Physics, Chinese Academy of Sciences, Beijing 100190, China}
\affiliation{School of Physical Sciences, UCAS, Beijing 100190, China}

\author{Hui Deng}
\affiliation{Synergetic Innovation Centre in Quantum Information and Quantum Physics,
USTC, Hefei 230026, China}

\author{Yi-Rong Jin}
\affiliation{Institute of
Physics, Chinese Academy of Sciences, Beijing 100190, China}
\affiliation{CAS Center for Excellence in Topological Quantum Computation, UCAS, Beijing 100190, China}

\author{Shu Chen}
\email{schen@iphy.ac.cn}
\affiliation{Institute of
Physics, Chinese Academy of Sciences, Beijing 100190, China}
\affiliation{School of Physical Sciences, UCAS, Beijing 100190, China}

\author{Dongning Zheng}
\email{dzheng@iphy.ac.cn}
\affiliation{Institute of
Physics, Chinese Academy of Sciences, Beijing 100190, China}
\affiliation{School of Physical Sciences, UCAS, Beijing 100190, China}
\affiliation{CAS Center for Excellence in Topological Quantum Computation, UCAS, Beijing 100190, China}
\affiliation{Songshan Lake Materials Laboratory, Dongguan 523808, China}

\author{Heng Fan}
\email{hfan@iphy.ac.cn}
\affiliation{Institute of
Physics, Chinese Academy of Sciences, Beijing 100190, China}
\affiliation{School of Physical Sciences, UCAS, Beijing 100190, China}
\affiliation{CAS Center for Excellence in Topological Quantum Computation, UCAS, Beijing 100190, China}
\affiliation{Songshan Lake Materials Laboratory, Dongguan 523808, China}

\date{ \today}

\begin{abstract}
A dynamical quantum phase transition can occur during time evolution of sudden quenched quantum systems across a phase transition.
It corresponds to the nonanalytic behavior at a critical time of the rate function of
the quantum state return amplitude, analogous to nonanalyticity of the
free energy density at the critical temperature in macroscopic systems. A variety of many-body systems
can be represented in momentum space as a spin-1/2 state evolving on the Bloch sphere,
where each momentum mode is decoupled and thus can be simulated independently
by a single qubit. Here, we report the observation of a dynamical quantum phase transition in a superconducting qubit simulation of
the quantum quench dynamics of many-body systems. We take the Ising model with a transverse field as an example for demonstration.
In our experiment, the spin state, which is initially polarized longitudinally, evolves
based on a Hamiltonian with adjustable parameters depending on the momentum and strength of the transverse magnetic field.
The time evolving quantum state is read out by state tomography. Evidence of
dynamical quantum phase transitions, such as paths of time evolution states on the Bloch sphere, non-analytic behavior
of the dynamical free energy and the emergence of Skyrmion lattice in momentum-time space, is observed.
The experimental data agrees well with theoretical and numerical calculations. The experiment demonstrates for the first
time explicitly the topological invariant, both topologically trivial and non-trivial, for dynamical quantum phase transitions.
Our results show that the quantum phase transitions of this class of many-body systems can be simulated successfully with a single
qubit by varying certain control parameters over the corresponding momentum range.
\end{abstract}

\maketitle

\section{Introduction}
Quantum simulation can provide insight into quantum and topological phases of matter, the role of
entanglement, and quantum dynamics \cite{timecrystal1,timecrystal2,JZ,Lukin,PJ,NF,MBL1,MBL2,MBL3,MBL4,MBL5,MBL6,YuYang1,YuYang2,GuoXY,sim1,sim2,sim3,sim4}.
It also constitutes one of the basic building blocks of quantum information
processors. Systems of high dimension or many-body systems can be simulated by the quantum processors with many coherently coupled
qubits \cite{JZ,Lukin}. By increasing the number of qubits,
the simulation may outperform classical machines, and demonstrate quantum advantage.
On the other hand, a variety of many-body systems with
a large number of spin-1/2 states can be studied in momentum space by a two-band model with decoupled
momentum modes, which is equivalent to a single spin-1/2 state evolving on the Bloch sphere
for each mode.
This fact also provides a route of quantum simulation with one qubit and variables sweeping over
momentum space, by means of coordinate momentum transformation.

In this Letter, we emulate the dynamical quantum phase transition (DQPT) of the many-body
systems by a single superconducting qubit. The DQPT is a phenomenon occurring in evolving
quantum states \cite{MH1,MH2,MH3} for isolated quantum systems far from equilibrium \cite{CG15}.
It is characterized by the non-analyticity in dynamical free energy
density at a critical time $t=t_c$, which is analogous to traditional phase transitions occurring at critical temperature.
The DQPT is intimately related to quantum phase transitions in many-body systems \cite{MH1,MH2,MH3,Chen,Lang,Ueda,Eck,MH4,AD,ZH,YL,JCB,CS,JG,MH5,Zvyagin}.

Recently, experimental explorations of DQPT have been performed in ion-trap systems \cite{PJ,JZ}
and cold atom systems \cite{NF,Lukin} with dozens of individual addressable qubits or a cloud of fermionic atoms.
Our experiment follows the DQPT simulation approach by emulating a corresponding two-band model
separately for each momentum mode with a single qubit. By ranging over the Brillouin zone of momentum space,
the results are equivalent to that of simulating many-body systems in space. The finite size effect can be observed
for a finite number of momenta implemented experimentally.
Our experimental system consists of superconducting Xmon qubits, which is one of the most promising platforms for quantum
simulation and quantum computation \cite{RBe,JKe,RB2e,NOe,ABe}. We provide concrete evidence that the DQPT is successfully simulated.
In particular,
we demonstrate experimentally the topological invariant in DQPT, which was studied recently in Refs.\cite{Chen,CS},
and have obtained quantitatively the dynamical free energy and Skyrmion lattice.

\section{The model and scheme for simulation}
We begin with a two-band model with Hamiltonian written in momentum space as
\begin{eqnarray}
{H}=\sum_{k}\Psi^\dagger_k h(k)\Psi_k ,
\end{eqnarray}
where $\Psi_k$ denotes a spinor, which is a 2-dimensional column vector formed by the fermion operators.
The ``first quantized'' Hamiltonian $h(k)$ for $k$ momentum mode takes the form
\begin{eqnarray}\label{hk}
h(k)=d_0(k)+{\bf d}(k)\cdot\boldsymbol{\sigma},
\label{hamilt}
\end{eqnarray}
where $\boldsymbol{\sigma}=(\sigma_x, \sigma_y, \sigma_z)$
is a vector of Pauli matrices and $k$ is in the Brillouin zone. This model can describe a variety of physically different many-body systems \cite{SSH,Kit,Barouch-Mccoy,Refael},
see appendix for details.

To study the quench dynamics, we first prepare the system in the ground state of the initial Hamiltonian $h_i(k)$, i.e.
$\rho _{i}(k)=|\phi _i(k)\rangle\langle\phi _i(k)|=\frac{1}{2}\left[1-\hat{\bf{d}} _i(k)\cdot\bf{\sigma}\right]$.
Then with a sudden quench to the final Hamiltonian $h_f(k)$,
which determines ${\bf d}_f(k)$ by Eq.(\ref{hamilt}),
the state evolves as
\begin{eqnarray}
\rho(k,t)=|\phi(k,t)\rangle\langle\phi(k,t)|=\frac{1}{2}\left[1-\hat{{\bf d}}(k,t)\cdot\bf{\sigma}\right],
\end{eqnarray}
where
\begin{eqnarray}
\hat{{\bf d}}\left(k,t\right)\cdot\boldsymbol{\sigma}=e^{-it{\bf d}_f\left(k\right)\cdot\boldsymbol{\sigma}}
\left(\hat{\bf{d}}_i\left(k\right)\cdot\boldsymbol{\sigma}\right)e^{it{\bf d}_f\left(k\right)\cdot\boldsymbol{\sigma}}.
\label{rotating}
\end{eqnarray}
This is simply the spin precession on the Bloch sphere, that is, $\hat{\bf{d}}_i(k)$ rotating around $\hat{\bf{d}}_f(k)$ with period $\frac{\pi}{|{\bf d}_f(k)|}$.

Now we introduce the rate function of the dynamical free energy,
\begin{eqnarray}
f(t)=-\frac{1}{N}\sum_k \log |\langle\phi_i(k)|e^{-ith_f(k)}|\phi_i(k)\rangle|^2.
\label{ratefunct}
\end{eqnarray}
The nonanalytic behavior of $f(t)$ corresponds to DQPT,
which is associated with zeros of $\langle\phi_i(k)|e^{-ith_f(k)}|\phi_i(k)\rangle$ for at least one critical momentum $k^*$ at critical time $t_c$.
From the spin precession picture, it is clear that spin vector ${\bf d}_i\left(k^*\right)$ is perpendicular to
the rotation axis ${\bf d}_f(k^*)$, where ${\bf d}_i\left(k^*\right)\neq 0$, ${\bf d}_f\left(k^*\right)\neq 0$,
and $t_c$ repeats with period $\frac{\pi}{|{\bf d}_f(k^*)|}$.

In this Letter, without loss of generality, we will investigate experimentally
the DQPT of the Ising model with a transverse field, but the approach
is applicable to other similar phenomena of many-body systems.
The Hamiltonian of the transverse field Ising model is
\begin{eqnarray}
H_{\rm Ising}=-\sum_{i=1}^{N}\left(\sigma^x_i\sigma^x_{i+1}+g\sigma^z_i\right),
\end{eqnarray}
where $g$ is the strength of the field in the $Z$ direction, and the periodic boundary
condition is assumed. There are two phases for this model,
the ferromagnetic phase for $g<1$, and the paramagnetic phase for $g>1$;
the phase transition critical point is $g_c=1$.
It is proved that DQPT occurs if and only if the initial Hamiltonian
with a $g_i$ field and the quenching Hamiltonian with a $g_f$ field belong to different phases \cite{MH1}.

The scheme for simulating DQPT in experiment is as follows. We first prepare the initial qubit state $|\phi _i(k)\rangle $ determined by
parameter $g_i$ for each mode $k$.
By the sudden quench, state $|\phi _i(k)\rangle $ evolves as $|\phi (k,t)\rangle $
according to the Hamiltonian, $h_f(k)=(g_f-\cos k)\sigma ^y+\sin k\sigma ^x$,
which depends on parameter $g_f$,
i.e., the spin vector $\hat{\bf{d}}_i(k)$ rotates around axis $\hat{\bf{d}}_f(k)$ on the Bloch sphere,
see appendix for details. The time
evolution state $|\phi (k,t)\rangle $ will be read out experimentally by state tomography.
By ranging over the Brillouin zone of momentum space for each mode $k$, we can obtain the rate function in Eq. (\ref{ratefunct}).
The occurrence of DQPT can be observed when
the rotation path of $|\phi (k,t)\rangle $ is a great circle on the Bloch sphere for mode $k=k^*$.
In this case, $|\phi (k^*,t)\rangle $ is orthogonal to the initial state at time $t_c$,
$\langle \phi _i(k^*)|\phi (k^*,t_c)\rangle =0$,
resulting in a nonanalytic point of the rate function. For the full regime of $k$ in the Brillouin zone,
the time evolutions of states $|\phi (k,t)\rangle $ will cover the full Bloch sphere when
there exists DQPT, otherwise only less than one half of the Bloch sphere
is covered, as recently pointed out by our co-authors \cite{Chen,CS}.
This phenomenon is observed, for the first time in experiment, as one of the signatures in identifying the occurrence
of DQPT. It is actually a direct observation of the topological invariant.

\begin{figure}[ht!]
%	\subfigure[ \label{Xmon_qubit}]
{\includegraphics[width=\linewidth]{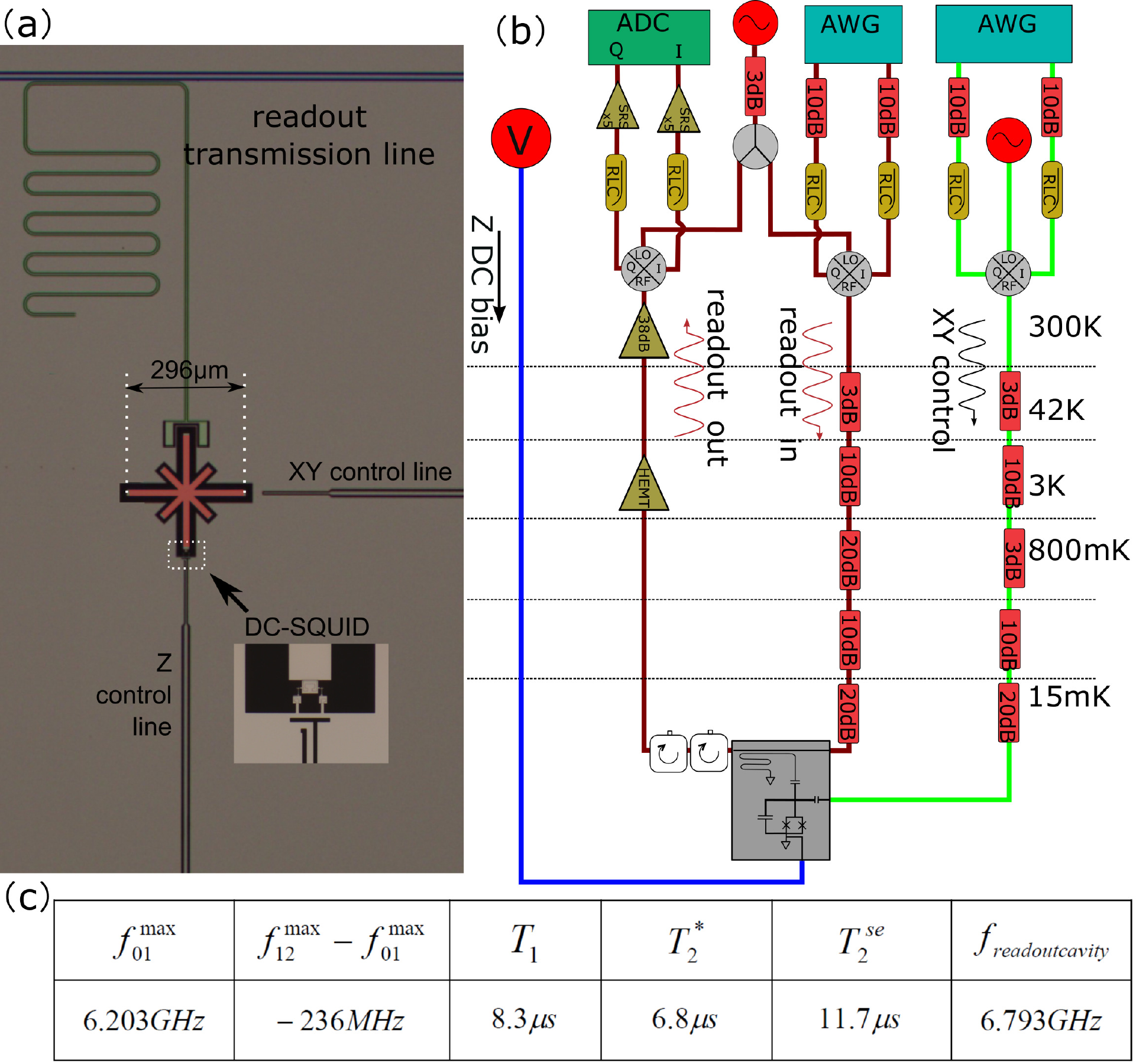}}	
%	\subfigure[ \label{Circuit digram}]{\includegraphics[scale=0.03]{circuitdigram.png}}
	\caption{Photography of qubit chip and external circuitries.
(a) is the microscopic photography of our Xmon qubit chip. The red part is the Xmon qubit.
Its frequency can be adjusted by applying DC current through its Z control line.
Transmission line coupled to the readout cavity is to measure the qubit state.
Basic information of qubit are listed in the table. The experiment data of the energy relaxation time $T_{1}$, dephasing time $T_{2*}$ and spin echo dephasing time $T_{2se}$
are also shown. (b) is the sketch of our experiment circuit setup, the blue part is for Z bias, the green part is for XY control, and the brown part is for readout.
(c) The qubit parameters are presented in the table.}\label{Fig1}
\end{figure}

\begin{figure*}[ht!]
  \centering
  \includegraphics[width=0.9\linewidth]{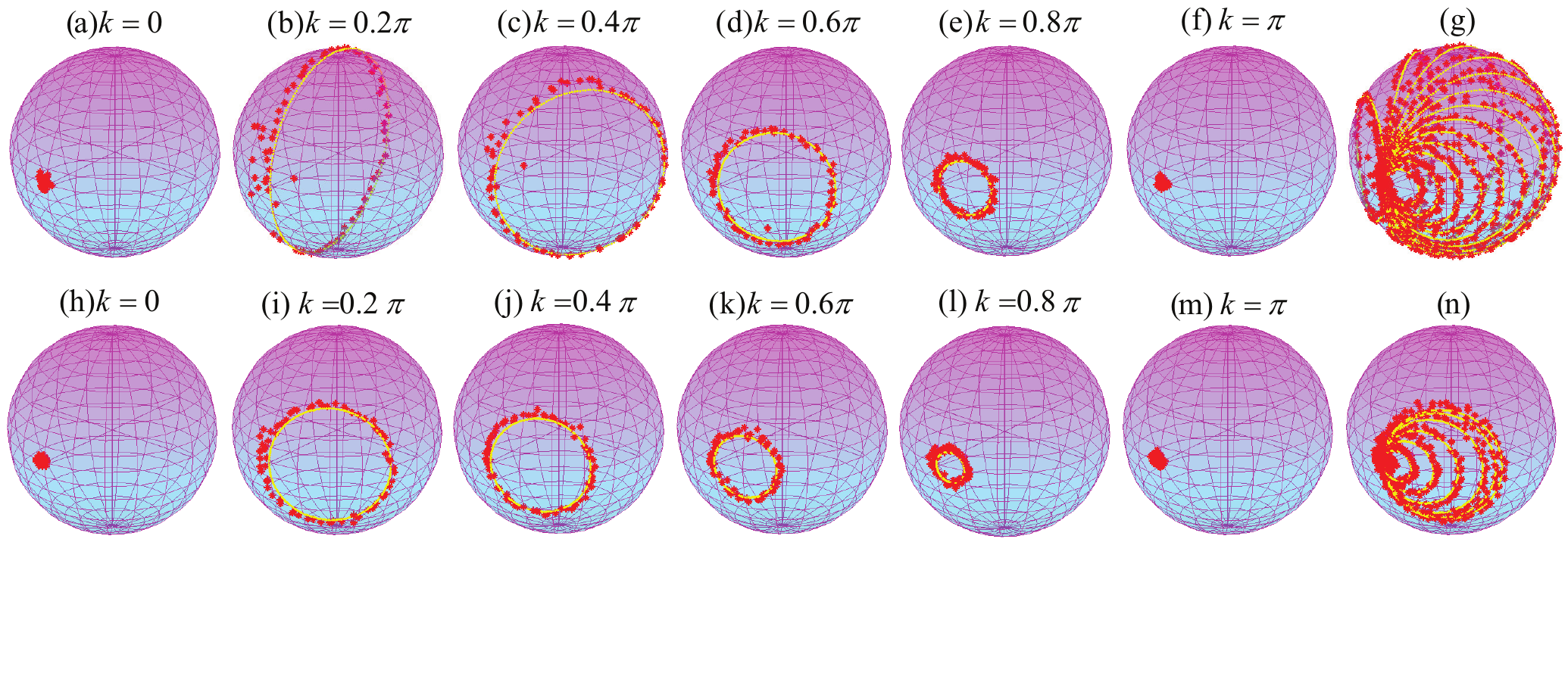}\\
  \caption{Evolution of states on the Bloch sphere. The state evolves depending on a fixed momentum.
  The data of evolving path are presented on the Bloch sphere.
  Here $g_i=0.2$ is fixed, cases with $g_f=1.5$ are presented in
  (a)-(g) in upper panel, cases of $g_f=0.5$ are presented in (h)-(n) in lower panel.
  The momenta are chosen to be $k=0, 0.2\pi , 0.4\pi , 0.6\pi , 0.8\pi, \pi$, presented respectively on up-down pairs of sub-figures, (a,h), (b,i), (c,j),(d,k), (e,l), (f,m).
  Data for two different cases are summarized together in (g) for $g_f=1.5$, and (n) for $g_f=1.5$, respectively.
  We can find that the whole Bloch sphere in (g) is covered, in contrast in (n), only partial region of the Bloch sphere is covered.
  We emphasize that the initial state is always prepared on the equator of the
Bloch sphere in X-axis.}\label{Fig2}
\end{figure*}

\section{Experimental setup}
In the experiment we use a single qubit to simulate the dynamics of the model.
Figure \ref{Fig1} is the microscopic photograph of the superconducting Xmon qubit chip \cite{RBe}, the external circuitries and
qubit parameters.
In experiment, the Xmon qubit is biased at its maximum frequency of $6.203$ GHz, --also known as the sweet-spot.
The measured anharmonicity is about $-236$ MHz,
the measured energy relaxation time $T_1$ about $8.3~\mu$s, dephasing time $T_{2*}$ about $6.8~\mu$s and spin echo dephasing time $T_{2se}$ about $11.7~\mu$s.
The readout cavity  frequency is about $6.793$ GHz, which falls in the dispersive coupling regime. The frequency dispersive shift of the readout cavity is $\kappa/2\pi = -0.697$ MHz.

The energy gap of the qubit can be adjusted by an external flux bias. The Xmon qubit is capacitively coupled to a $\lambda / 4$ coplanar waveguide (CPW) resonator that is coupled to a CPW transmission line. In this device, the qubit state is read out by the dispersive method via the $\lambda / 4$ resonator.  The optical micrograph of this sample is shown in Fig.~\ref{Fig1}(a).
The details of chip fabrication and the circuitry are presented in appendix.

\section{Time evolution paths on the Bloch sphere for DQPT}
Following our experimental scheme, we first
prepare the initial state as the ground state of the Hamiltonian $h_i(k)$ for a fixed mode $k$, then suddenly
quench the system to the final Hamiltonian $h_f(k)$.
For convenience, we actually always prepare the initial state as $|\phi _i\rangle =(|0\rangle +|1\rangle )/\sqrt{2}$,
consequently the quenched Hamiltonian is changed accordingly. This is because we can perform a rotation
to both Hamiltonians, $h_i(k)$ and $h_f(k)$, without changing the DQPT results.

The quenched quantum state will be read out at a sequence of time points to obtain
the time dependent density matrix $\rho (k,t)$. For a full rotation period, we
can obtain a circular evolution path of the state on the Bloch sphere. The same procedure repeats by changing momentum $k$
in the Brillouin zone.

In the experiment, we let $g_i=0.2$, which is in the ferromagnetic phase regime.
The system is suddenly quenched to the final Hamiltonian $h_f(k)$. Here
two different strengths of the field are chosen, $g_f=0.5$ and $g_f=1.5$, corresponding to
the ferromagnetic and
paramagnetic phases, respectively.

The qubit is first rotated about the Y-axis by a $\pi /2$ microwave pulse to the superposed state
$|\phi _i\rangle =(|0\rangle +|1\rangle )/\sqrt {2}$. For a fixed k, a unitary operation based
on the final Hamiltonian is applied to the initial state $|\phi _i\rangle $ as the quantum quench procedure.
We then sweep mode $k$ in the Brillouin zone from 0 to $2\pi $ with step length $2\pi /30$.

For each value of $k$, the state will be rotated for two cycles on the Bloch sphere,
representing time evolution for two periods. The rotation axis is determined by the quench Hamiltonian $h_f(k)$.
The path of the state time evolution is presented in Fig.~\ref{Fig2}, where only one cycle
of data is presented. In the figure, each dot represents the evolving state at a fixed time point
read out experimentally by state tomography. For example, Figs.~\ref{Fig2}(a) and (h) represent $k=0$ for different quenched Hamiltonians. We can find that
the initial state always stays at its original position, because
the rotation axis is the X-direction determined by the corresponding Hamiltonian.

Figure~\ref{Fig2}(a-g), 7 sub-figures in the upper panel, represent the system is suddenly quenched to $g_f=1.5$,
and the state evolutions on the Bloch sphere for $k=0, 0.2\pi , 0.4\pi , 0.6\pi, 0.8\pi , \pi $ are presented in the first 6 sub-figures, respectively.
All data for this case are presented together in Fig.~\ref{Fig2}(g),
where those $k$ modes are for $k\in [0,\pi ]$ constituting a half region in the Brillouin zone.
For each mode $k$, the state starts from $|\phi _i\rangle $ in the original position and evolves like
a circle on the Bloch sphere.
In each cycle of time period, we take 70 time points for state tomography readout. The experimental data
are presented as dots on the Bloch sphere, where each dot represents average value of
5000 single-shot measurement results.
Each step of time evolution lasts 15 nanoseconds.
Then one circle of period takes 1.05 $\mu $s,
two circles are also performed experimentally, they are within the coherence time.
We have also taken a normalization, $|<|\phi(k,t)|\phi(k,t)>|^2=1$, at each time point, implying
pure states are assumed for time evolution, ${\rm Tr}\rho ^2(k,t)=1$.
We take total $30$ different momenta $k$ in the Brillouin zone in experiment,
the evolution paths are given in Fig.~\ref{Fig2}(g).

Figure~\ref{Fig2}(h-n), 7 sub-figures in the lower panel, represent the case that
the system is suddenly quenched to $g_f=0.5$. Similar conventions are used as those in upper panel.

The occurrence of DQPT can be directly observed in Fig.~\ref{Fig2}. It is obvious that in upper panel
of the figure, Fig.~\ref{Fig2}(a-g), the full Bloch sphere is covered by states time evolution paths
shown explicitly in Fig.~\ref{Fig2}(g).
This case is that $g_f=1.5$ and $g_i=0.2$ are located in two different phases, so DQPT happens.
Since the full Bloch sphere is covered, it is apparent that
there exists a $k^*$, the path of the evolving state is a great circle resulting in that state $|\phi ^{\perp }_i\rangle =(|0\rangle -|1\rangle )/\sqrt {2}$,
located in the opposite direction of X-axis on the Bloch sphere, can be reached at a critical time $t_c$,
shown in Fig.~\ref{Fig2}(b). The orthogonality leads to zero for
overlap between the evolving state $|\phi ^{\perp }_i\rangle $ with the initial state $|\phi _i\rangle $, leading to non-analyticity
for logarithm in the rate function (\ref{ratefunct}). These results demonstrate the occurrence of DQPT at a critical time $t_c$.
In contrast, when $g_i=0.2$ is quenched to $g_f=0.5$ but without going across the
critical point $g_c=1$, we can observe in Fig.~\ref{Fig2}(h-n) that only less than one half of the Bloch sphere
is covered for $k$ in the Brillouin zone, as summarized in Fig.~\ref{Fig2}(n). Then no DQPT can happen.

\begin{figure}[ht!]
  \centering
  \includegraphics[width=0.9\linewidth]{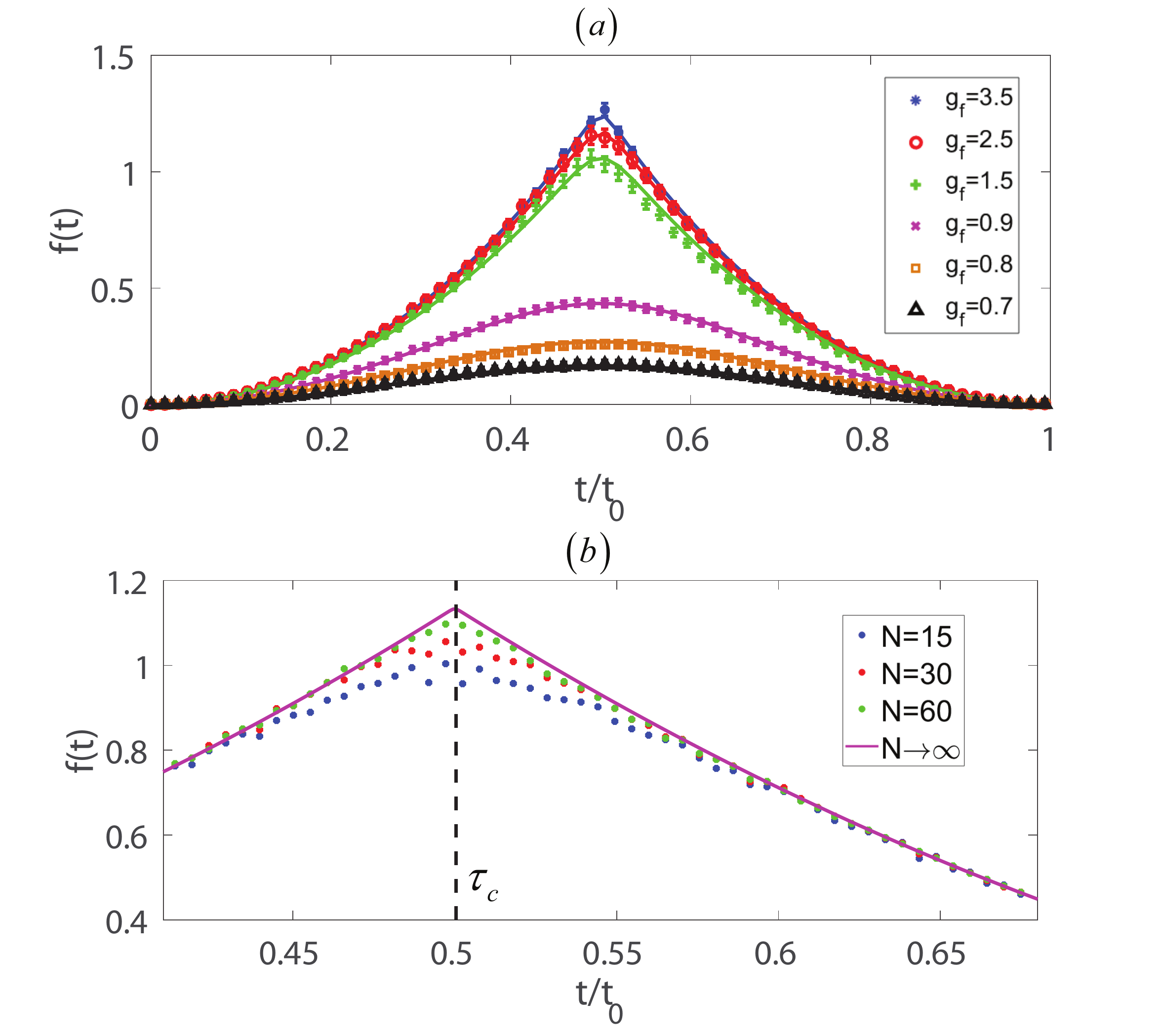}\\
  \caption{The dynamical free energy. (a) Dynamical free energies for different $g_f$s are presented,  while $g_i=0.2$ is fixed.
  We take $70$ time points for each period. Error bar represents the deviation
  of the average value of 5000 single-shot measurements from the fitting value
  for the evolution path of 70 points on the Bloch sphere, see appendix
  for details.
  (b) The dynamical free energies near the critical time $\tau_c$ for different
  number of momenta implemented in experiment are given, corresponding to different sizes.
  The exact results for $N\rightarrow \infty$ are presented as solid line. Here we take $g_i=0.2$ and $g_f=1.5$.}\label{Fig3}
\end{figure}

\section{The Rate function, finite size effect and the Skyrmion lattice}
Quantitatively, we can obtain the evolution of dynamical free energy defined in Eq.(\ref{ratefunct}).
Figure~\ref{Fig3}(a) presents the time dependent rate functions for different $g_f$, all with initial parameter $g_i=0.2$.
The experimental data are shown as dots, the theoretical results are presented as lines.
We can find that the rate functions $f(t)$ have sharp peaks at the critical time $t_c$ for $g_f=3.5, 2.5, 1.5$, which lead
to discontinuity for derivative of $f(t)$ at $t_c$. This phenomenon corresponds to the DQPT.
In comparison, it is obvious that the rate functions for $g_f=0.9,0.8,0.7$ are different from cases when $g_f>1$.
The curves are much more smooth and no sharp peak appears, so no discontinuity is expected for derivative of the rate functions.
Thus no DQPT will happen. The results agree well with theoretical calculations.

Figure~\ref{Fig3}(b) shows the results of different number $N$ of modes for $k$, corresponding
to size $N$ of the Ising model. So experiments are performed for $N$ equally separated momenta for $k\in [0,2\pi]$.
Here $g_i=0.2$ and $g_f=1.5$ are fixed.
It can be found that if $t$ is far away from $t_c$, the dynamical free energy $f(t)$ is quite close to theoretic value (pink curve) of $N\rightarrow\infty$.
Near the critical time $t_c$, $f(t )$ is nearly smooth if the size is small, demonstrating finite size effect.
As $N$ increases, it approaches to the theoretical value for $N\rightarrow \infty$ and demonstrates non-analytic behavior.

\begin{figure}[htbp]
  \centering
  \includegraphics[width=1\linewidth]{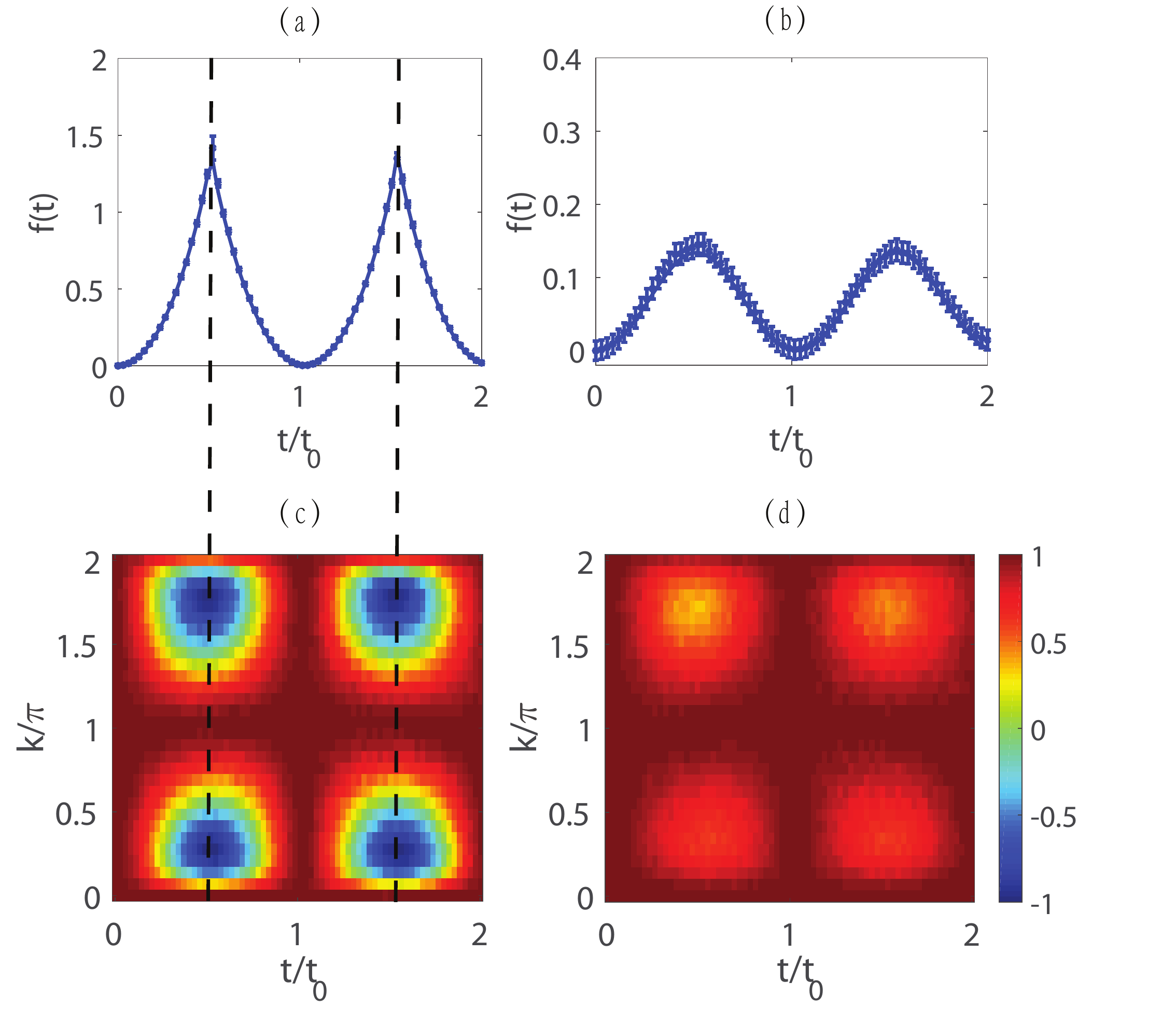}\\
  \caption{Skyrmion and DQPT. (a) and (b) are respectively dynamical free energies for $g_f=1.5$ and $g_f=0.5$, both
  with $g_i=0$. The number of momenta is $N=30$. For each momentum mode $k$, states at $2\times 70=140$ time points are read out.
  Error bar is deviation of the average value from the fitting value, see appendix for details.
  (c) and (d) are expectation values of the initial spin operator about the evolving state $|\phi(k,t)>$.
  The Skyrmion are shown obviously in (c), while no Skyrmion appears in (d).}\label{Fig4}
\end{figure}

Figure \ref{Fig4} shows the emergence of Skyrmion lattice in momentum-time space for DQPT.
We define the expectation value as,
$\langle \hat{\bf d}(k,t)\rangle =\langle \phi (k,t)|\hat{\bf d}_i|\phi (k,t)\rangle $, see
also appendix.
We consider two different cases, $g_f=1.5$ in Fig. \ref{Fig4}(c) and $g_f=0.5$ in Fig. \ref{Fig4}(d),
both start from the initial condition $g_i=0$. The rate functions are presented respectively on Fig. \ref{Fig4}(a,b) for comparison.
We find that when $g_i=0$ and $g_f=1.5$ lie in different phases,
the emergence of Skyrmion lattice in momentum-time space can be seen obviously in Fig. \ref{Fig4}(c),
which indicates the nontrivial dynamical Chern number implying the occurrence of DQPT.
The time coordinates of the center of Skyrmion is just the critical time $t_c$.
While if $g_i=0$ and $g_f=0.5$ lie in the same phase, the configuration of Skyrmion lattice does not appear as shown in Fig. \ref{Fig4}(d),
the corresponding dynamical Chern number is trivial. There is no DQPT as shown Fig. \ref{Fig4}(b).

\section{Conclusion and Discussion}
In summary, we simulate successfully the two-band model of DQPT
 for the transverse field Ising model by a single superconducting qubit. The DQPT is
 shown by state evolution paths on the Bloch sphere, the dynamical free energy and the Skyrmion lattice.
 The critical time of DQPT is quantitatively identified.
 This approach is applicable in investigating various physical phenomena of the class of free fermionic many-body systems.
The similar experimental scheme can be applied to simulate temporal topological phenomena
by demonstrating that a single qubit is driven by two elliptically polarized periodic waves \cite{Refael}.

In our scheme, phenomena of many-body system are simulated by a single qubit at the expenses of 
repeating experiments by ranging over the momentum space. On the other hand, besides the two-level system, 
it is known that the superconducting Josephson junction can have controllable multiple energy levels.
Then, this platform is promising for more simulating applications, such as the PT-symmetric physics, 
geometric quantum logic gates for quantum computation. Also, the superconducting qubit or multi-level system can 
be coupled to bosonic modes by a resonator or cavity, simulations such as 
spin-boson phenomena, quantum random walks and quantum statistical models are expected.
So, our results pave the way for more applications of the superconducting quantum platform.

\begin{acknowledgments}
The first two authors, X.Y.G. and C.Y., contributed equally to this work.
We thank Wuxin Liu and Haohua Wang of Zhejiang University for technical support.
We thank Ling-An Wu for careful reading our manuscript to improve our presentation.
	This work was supported by National Key Research and Development Program of China (Grant Nos. 2016YFA0302104, 2016YFA0300600, 2014CB921401, 2017YFA0304300),
	National Natural Science Foundation of China (Grant Nos. 11425419, 11404386, 11674376, 11774406),
and Strategic Priority Research Program of Chinese Academy of Sciences (Grant No. XDB28000000).
\end{acknowledgments}

\section*{Appendix A: The qubit device and external circuitries}

The sample was fabricated using a process involving electron-beam-lithography (EBL) and double-angle evaporation. In brief, a 100 nm thick Al layer was firstly deposited on a $10 \times 10$ mm sapphire substrate by means of electron-beam evaporation, followed by EBL and wet etching to produce large structures such as microwave coplanar-waveguide resonators/transmission lines, capacitors of Xmon qubit and electric leads. The EPL resist used was ZEP520 and wet etching process was carried out using Aluminum Etchant Type A. In the next step, the Josephson junctions of qubits were fabricated using the double-angle evaporation process. In this step, the under cut structure was created using a PMMA-MMA double layer EBL resist following a process similar to that reported in Ref.~\cite{RBe}. During the evaporation, the bottom electrode was about $30$ nm thick while the top electrode was about $100$ nm thick with intermediate oxidation.

In the measurements, the sample was mounted in an aluminum alloy sample box which is fixed  on the mixing chamber stage of a dilution refrigerator. The temperature of the mixing chamber was below 15 mK during measurements. The readout input microwave lines and qubit XY control lines are heavily attenuated. Lines for qubit dc bias control are filtered using filters (RLC ELECTRONICS F-10-200-R) that functions as combination of low-pass filter and copper powder filter.  The microwave output signal from the transmission line is amplified ($\approx 39 ~\mathrm{dB}$) by a cryogenic HEMT amplifier mounted at the 4 K stage and a room temperature amplifier ($\approx 38 ~\mathrm{dB}$) before being measured by a home-built heterodyne acquisition system shown in Fig.~1(b) in the main text.

\section*{Appendix B: Physical description of the superconducting qubit}
\begin{figure*}[!ht]
			\centering
			\subfigure[Isolated transmon circuit\label{iso_transmon}]
			{\includegraphics[height=0.15\textwidth]{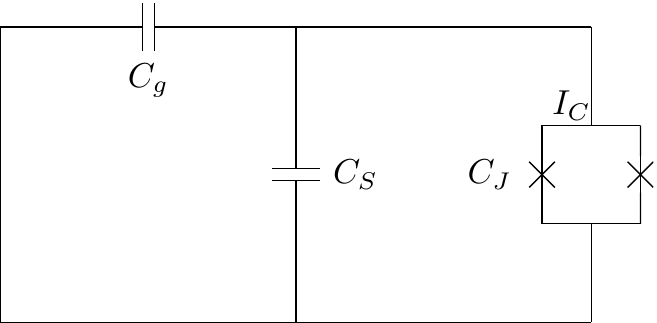}}
\hskip 1truecm
			\subfigure[Xmon circuit with control\label{ctrl_xmon}]{\includegraphics[height=0.15\textwidth]{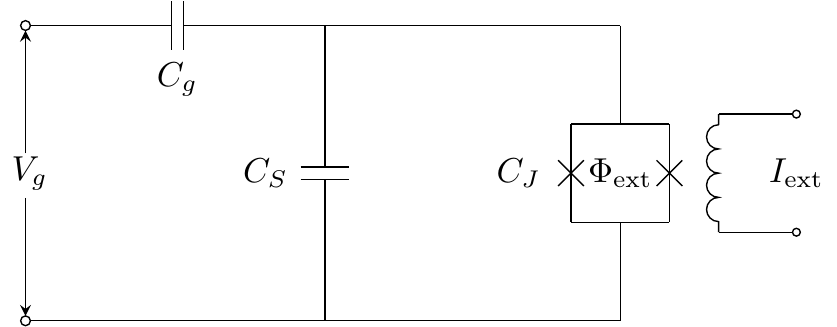}}
			\caption{Equivalent circuits of (a) the isolated transmon circuit and (b) its Xmon variety with control.}
\label{Xmon}
		\end{figure*}

		In the past decades, there has been a great progress in the field of superconducting qubits.  The main aims
		are to achieve better control and longer coherent time for qubit or qubits. As a result, many types of superconducting
		qubits have been developed, each of which has its own advantages and limits. There are three main categories of the quantum
		superconducting qubit working in different regimes according the ratio of the Josephson energy $E_J$ to the
		charging energy $E_C$~\cite{Fink2010,Girvin2011,Wendin2017,Gu2017}: 1) charge qubit with $E_J/E_C\sim0.1$; 2) flux qubit with $E_J/E_C\sim50$;
		and 3) phase qubit
		$E_J/E_C\sim10^6$.
		%Despite the different design, the central part of a superconducting qubits
		By adding a large capacitor $C_S$ parallel to the superconducting quantum interference device (SQUID) and thus
		shunting the later (cf. Fig.~\ref{iso_transmon}), the transmon qubit works in the parameter regime of  $E_J/E_C$
		being the order of several tens or several hundreds.
		It gains the advantage of exponentially suppressing the sensitivity to the charge noise at the expense of
		polynomial reduction of the anharmonicity~\cite{JKe,Fink2010,Girvin2011,Wendin2017,Gu2017}. Notice that
		anharmonicity describes the variation of the energy level spacing which ensures the possibility of addressing the
		lowest energy levels of the platform. The Hamiltonian of an isolated transmon qubit is
		\begin{equation}
			\hat{H}_0 = 4E_C\hat{n}^2 - E_J\cos\hat{\phi}.
			\label{iso_Hamiltonian}
		\end{equation}
		where $\hat{n}$ is the operator corresponding to the number of Cooper pair tunneled through the Josephson junctions and
		$\hat{\phi}$ denotes the gauge-invariant phase difference operator across the Josephson junctions. They are
		mutually conjugate and satisfy the commutation relation $\lbrack{\hat{\phi},\hat{n}}\rbrack=i$.
		The charging energy $E_C=e^2/2C_\Sigma$ depends on the total capacitance $C_\Sigma=C_S+C_g+C_J$ of the shunt capacitor
		$C_S$, gate capacitor $C_g$ and the Josephson junction capacitance $C_J$. The Josephson energy
		$E_J={\hbar{I_C}}/{2e}$ is determined by the critical current of the DC-SQUID,
which is modulated by the external magnetic flux.
		In the transmon regime, $C_S$ is very large
		such that $20{\lesssim}E_J/E_C{\lesssim}100$, its  $\ell$-th eigenenergy
		level should be~\cite{JKe,Fink2010,Girvin2011,Wendin2017,Gu2017}
		\begin{equation}
			E_\ell \simeq -E_J + \sqrt{8E_JE_C}(\ell+1/2) - \frac{1}{2}E_C\lbrack{\ell(\ell+1)+1/2}\rbrack.
		\end{equation}
		The anharmonicity is big enough for the addressability of the two lowest energy levels and thus constitutes a qubit
		\begin{equation}
			\hat{H}_0 = \frac{1}{2}\hbar\omega\hat{\sigma}_z
			\label{iso_Hamiltonian_tls}
		\end{equation}
		with $\omega=2(\sqrt{8E_JE_C}-E_C)/\hbar$ and $\hat{\sigma}_z=\ket{e}\bra{e}-\ket{g}\bra{g}$. Here $\ket{g}$ is the
		ground state of the transmon while $\ket{e}$ is its first excited eigenstate.

		The quantum platform we employed is a superconducting Xmon
		qubit, which is designed on the basis of coplanar transmon.
		Essentially it is equivalent to a grounded transmon, see Fig.~\ref{Xmon}. Embedded in an uninterrupted ground plane, the Xmon qubit
		can prolong the coherent time by further employing coplanar waveguide made with high-quality material. Better connectivity
		can be accomplished via a cross-shaped capacitor~\cite{RBe,Fink2010,Girvin2011,Wendin2017,Gu2017}.
		The driving microwave applied as shown in Fig.~\ref{ctrl_xmon} is the following
		\begin{equation}
			V_g = V\cos(\omega{t}+\phi_0).
		\end{equation}
		The frequency of the microwave is chosen to match the resonant frequency of the isolated qubit in
		Eq.(\ref{iso_Hamiltonian_tls}). As a result, the Hamiltonian of the Xmon is
		\begin{eqnarray}
			\hat{H}
			&=& 4E_C\hat{n}^2 + \frac{V_gC_g\hat{n}e}{C_\Sigma} - E_J\cos\hat{\phi}.
		\end{eqnarray}
		The two-level qubit Hamiltonian via truncating all the higher energy levels is thus
		\begin{equation}
			\hat{H} = \frac{1}{2}\hbar\left(\omega+\omega_\Phi\right)\hat{\sigma}_z
			+ \frac{AC_ge}{C_\Sigma}\cos(\omega{t}+\phi_0)\hat{\sigma}_x,
		\end{equation}
		where $\omega_\Phi$ is the energy level shift caused by the external magnetic flux $\Phi_\mathrm{ext}$ controlled by
		varying $I_\mathrm{ext}$. Moving to the interaction picture with respect to $\hat{H}_0$, we would have
		\begin{equation}
			\hat{H}_r
			= \hbar\omega_\Phi\hat{\sigma}_z
			+\frac{AC_ge}{C_\Sigma}\left(\cos\phi_0\hat{\sigma}_x+\sin\phi_0\hat{\sigma}_y\right).
\label{controlH}
		\end{equation}
		By varying the amplitude $A$ and phase $\phi_0$ of the driving voltage $V_g$ which is applied through the gate capacitor
		$C_g$, we would have full control of the rotations of the qubit along the X as well as the Y direction. The
		Z-direction control is exerted via the change control current $I_\mathrm{ext}$ which adjusts the external magnetic flux
		$\Phi_\mathrm{ext}$ thrusting through the SQUID loop.

\section*{Appendix C: The many-body systems and the two-band model}

The Hamiltonian of a two-band model is written as,
\begin{eqnarray}
{H}=\sum_{k}\Psi^\dagger_k h(k)\Psi_k
\label{s-Hamiltonian}
\end{eqnarray}
where $\Psi_k$ denotes a spinor, $h(k)$ takes the form
\begin{eqnarray}\label{hk}
h(k)=d_0(k)+{\bf d}(k)\cdot \boldsymbol{\sigma},
\end{eqnarray}
where $\boldsymbol{\sigma}=(\sigma_x, \sigma_y, \sigma_z)$
is a vector of Pauli matrices, as already presented in the main text.
This model can describe a variety of physically different many-body systems.
For examples, the Su-Schrieffer-Heeger (SSH) model \cite{SSH} describes the simplest one-dimensional topological insulator.
We have that $\Psi^\dagger_k=(c^\dagger_{kA},c^\dagger_{kB})$ with A and B referring two sub-lattices,
${\bf d}(k)=[(t+\delta t)+(t-\delta t)\cos k, (t-\delta t)\sin k, 0)]$, with $(t\pm \delta t)$ being the hopping amplitudes in the unit cell
and between the adjacent cells, respectively. Another example is p-wave Kitaev chain \cite{Kit} which
describes a one-dimensional topological superconductor. For this case, we have that $\Psi^\dagger_k=(c^\dagger_{k},c_{-k})$
and ${\bf d}(k)=[0, \Delta \sin k, (-\mu/2-t\cos k)]$, where $\Delta$ denotes the pairing potential, $\mu$ the chemistry potential, $t$ the hopping amplitude.
When $t=\Delta$, it corresponds to the transverse field Ising model \cite{Barouch-Mccoy} after mapping to the free fermions by Jordan-Wigner transformation.
Here, we have ${\bf d}(k)=[0, \sin k, (g-\cos k)]$. The details are as follows.

The SSH model is the simplest two-band model describing polyacetylene, which is a one-dimensional topological insulator. The Hamiltonian reads
\begin{eqnarray}
H_{SSH}=\sum_{l}(t+\delta t)c^\dagger_{A,l}c_{B,l}+(t-\delta t)c^\dagger_{A,l+1}c_{B,l}+ h.c.
\nonumber \\
\end{eqnarray}
A and B refer to two sublattices. The hopping amplitude in the unit cell is $t+\delta t$ while that between adjacent unit cell is $t-\delta t$. Performing the Fourier transformation $c_{kA}=\frac{1}{\sqrt{N}}\sum_l e^{-ikl}c_{A,l}$ and $c_{kB}=\frac{1}{\sqrt{N}}\sum_l e^{-ikl}c_{B,l}$, where $N$ is the number of sites, we obtain
\begin{eqnarray}
H_{SSH}=&\sum_{k}&(t+\delta t)(c^\dagger_{kA}c_{kB}+h.c.)\nonumber\\&+&(t-\delta t)(e^{ik}c^{\dagger}_{kA}c_{kB}+h.c.).
\end{eqnarray}
Introducing the spinor $\Psi ^{\dagger}_{k}=(c^\dagger_{kA}, c^\dagger_{kB})$, the Hamiltonian can be written in a compact form,
\begin{eqnarray}
H_{SSH}=&\sum_{k}&\Psi^\dagger_k[((t+\delta t)+(t-\delta t)\cos k)\sigma_x\nonumber\\
&+&(t-\delta t)\sin k\sigma_y]\Psi_k,
\end{eqnarray}
where ${\bf d}(k)=[(t+\delta t)+(t-\delta t)\cos k, (t-\delta t)\sin k, 0)]$ referring to Eq.(\ref{hk}). The system is topologically nontrivial when $|t+\delta t|<|t-\delta t|$. Otherwise it is topologically trivial.

The p-wave Kitaev chain is a one-dimensional topological superconductor introduced by Kitaev \cite{Kit}. The Hamiltonian reads,
\begin{eqnarray}
H_{K}=\sum_{l}&-&t(c^\dagger_lc_{l+1}+h.c.)-\Delta(c^\dagger_lc^\dagger_{l+1}+h.c.)\nonumber\\
&-&\mu(c^\dagger_lc_l-\frac{1}{2})
\end{eqnarray}
$t$ is the hopping amplitude; $\Delta$ is the p-wave superconductor pairing potential and $\mu$ is the chemical potential. Performing the Fourier transformation $c_{k}=\frac{1}{\sqrt{N}}\sum_l e^{-ikl}c_{l}$ and introducing the spinor $\Psi^\dagger_k=(c^\dagger_k, c_{-k})$, we obtain,
\begin{eqnarray}
H_K=\sum_{k}\Psi^\dagger_k[\Delta \sin k\sigma_y+(-\mu/2-t\cos k)\sigma_z]\Psi_k,
\end{eqnarray}
where ${\bf d}(k)=[0, \Delta \sin k, (-\mu/2-t\cos k)]$ referring to Eq.(\ref{hk}). This model is mathematically equivalent to the transverse field Ising model when $t=\Delta$, and it is topologically nontrivial when $\frac{2|t|}{|\mu|}>1$.

The transverse field Ising model is described as,
\begin{eqnarray}
H_{Ising}=-\sum_{l}\sigma^x_l\sigma^x_{l+1}+g\sigma^z_l.
\label{SIsing}
\end{eqnarray}
$g$ is the transverse field strength. The spin model can be mapped to the free-fermion model by using Jordan-Wigner transformation
\begin{eqnarray}
\sigma^z_l=1-2c^\dagger_lc_l,~~~\sigma^x_l=\prod_{j<l}(1-2c^\dagger_jc_j)(c_l+c^\dagger_l).
\end{eqnarray}
The Hamiltonian changes to,
\begin{eqnarray}
H_{Ising}=-\sum_l(c^\dagger_lc_{l+1}+c^\dagger_lc^\dagger_{l+1}+h.c.)+g(1-2c^\dagger_lc_l).
\nonumber \\
\end{eqnarray}
Again, by using Fourier transformation and introducing the spinor $\Psi^\dagger_k=(c^\dagger_k,c_{-k}))$, we obtain,
\begin{eqnarray}
H_{Ising}=\sum_{k}\Psi^\dagger_k[\sin k\sigma_y+(g-\cos k)\sigma_z]\Psi_k,
\label{SIsingeff}
\end{eqnarray}
where ${\bf d}(k)=[0, \sin k, (g-\cos k)]$, which is used in the main text.
It is well known that the model is in ferromagnetic phase when $g<1$ and in paramagnetic phase when $g>1$.

For Hamiltonian (\ref{s-Hamiltonian}), see also (1) in the main text,
one can find that each k mode is decoupled, so we can investigate each mode separately. The eigenvalues of $h(k)$ are given by
\begin{eqnarray}
\epsilon_{\pm}(k)=d_0(k)\pm |{\bf d}(k)|.
\end{eqnarray}
The corresponding eigenvectors are denoted by $|\phi_{\pm}(k)\rangle$, or written as density matrices
\begin{eqnarray}
\rho_{\pm}(k)=|\phi_{\pm}(k)\rangle\langle\phi_{\pm}(k)|=\frac{1}{2}\left[1\pm\hat{\bf{d}}(k)\cdot\boldsymbol{\sigma}\right],
\label{minuseigen}
\end{eqnarray}
where $\hat{\bf{d}}(k)=\frac{{\bf d}(k)}{|{\bf d}(k)|}$ corresponding to a unique vector on the Bloch sphere.

To study the quench dynamics, we first prepare the system in ground state of the initial Hamiltonian $h_i(k)$, i.e.
$\rho _{i}(k)=|\phi _i(k)\rangle\langle\phi _i(k)|=\frac{1}{2}\left[1-\hat{\bf{d}} _i(k)\cdot\bf{\sigma}\right]$, corresponding
to the minus eigenvector in (\ref{minuseigen}). Then taking a sudden quench to the final Hamiltonian $h_f(k)$,
which determines ${\bf d}_f(k)$.
The state evolves as $|\phi(k,t)\rangle=e^{-ith_{f}(k)}|\phi_i(k)\rangle$.
A more enlightening picture can be presented as density matrix form,
\begin{eqnarray}
\rho(k,t)=|\phi(k,t)\rangle\langle\phi(k,t)|=\frac{1}{2}\left[1-\hat{{\bf d}}(k,t)\cdot\bf{\sigma}\right],
\end{eqnarray}
where
\begin{eqnarray}
\hat{{\bf d}}\left(k,t\right)\cdot\boldsymbol{\sigma}=e^{-it{\bf d}_f\left(k\right)\cdot\boldsymbol{\sigma}}
\left(\hat{\bf{d}}_i\left(k\right)\cdot\boldsymbol{\sigma}\right)e^{it{\bf d}_f\left(k\right)\cdot\boldsymbol{\sigma}}.
\label{rotating}
\end{eqnarray}
It is simply the spin precession on the Bloch sphere, that is, $\hat{\bf{d}}_i(k)$ rotates around $\hat{\bf{d}}_f(k)$ with period $\frac{\pi}{|{\bf d}_f(k)|}$.

\begin{figure*}[!ht]
			\centering
			{\includegraphics[height=0.215\textwidth]{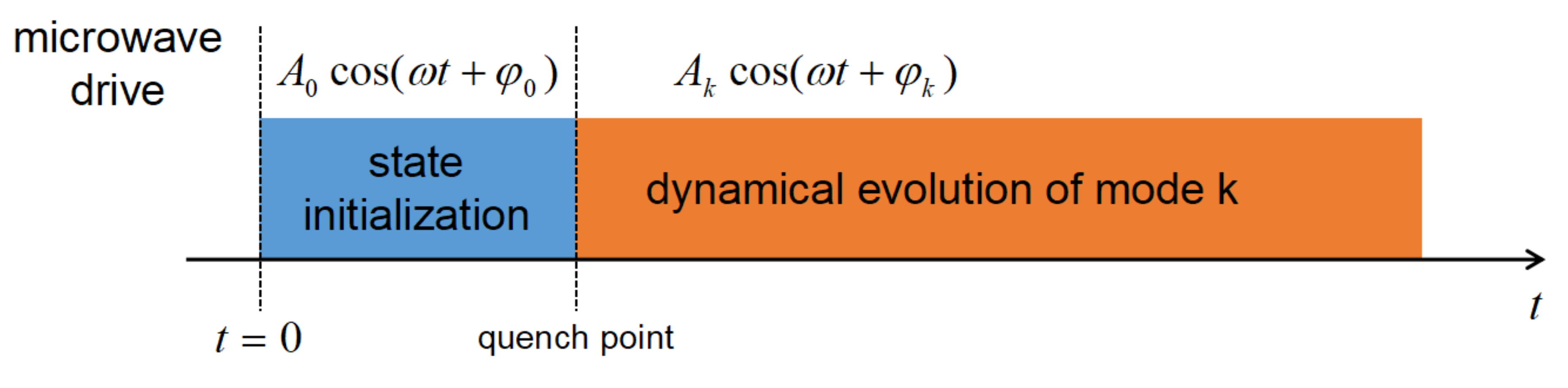}}
			\caption{Experiment control sequence. The initial state is prepared at the state initialization period by control quantity
$A_0\cos (\omega t+\varphi _0)$. The state is prepared as $|\phi _i\rangle $. Then for a quantum quench, by controlling
$\varphi _k$ depending on momentum $k$, we adjust the direction of the rotation axis, shown in Eq.(\ref{controlH}).}\label{FigS2}
\end{figure*}

\section*{Appendix D: Experimental scheme}
Experimentally, we prepare the initial state and control its evolution by the corresponding Hamiltonian,
see Fig.~\ref{FigS2} for the schematic description. The evolving state is read out by state tomography.
Our simulation focuses on the case of the transverse field Ising model (\ref{SIsing},\ref{SIsingeff}). In general, the initial state should be prepared as $|\phi _i(k)\rangle $ depending on the
initial Hamiltonian $h_i(k)$. Without loss of generality, we always prepare the initial state in experiment as
\begin{eqnarray}
|\phi _i\rangle =\frac {1}{\sqrt {2}}(|0\rangle +|1\rangle ).
\label{Sintial}
\end{eqnarray}
At the same time, the quenched Hamiltonian should be changed correspondingly. Considering that the prepared initial state $|\phi _i\rangle $ in Eq.(\ref{Sintial})
is an eigenvector of $\sigma ^x$, $\sigma ^x=U^{\dagger }h_i(k)U$, which corresponds to $h_i(k)$ by a unitary transformation $U$.
Then for a quantum quench, the applied Hamiltonian takes the form $U^{\dagger }h_f(k)U$.

Now, let us show how to realize a qubit rotation in experiment. In the Hamiltonian of a qubit Eq.(\ref{controlH}), the two terms in the parentheses
represent a rotation with axis in the XY plane. The direction can be adjusted by controlling the parameter $\phi _0$ in Eq.(\ref{controlH}),
which depends on momentum $k$. Experimentally as shown in Fig.~\ref{FigS2}, by controlling $\varphi _k$ in $A_k\cos (\omega t+\varphi _k)$,
we can realize the control of axis direction in the XY plane for a rotation. Explicitly in Fig.2 in the main text, we can find that the rotation axes are in
the XY plane.

Our simulation scheme can be applied to general two-band models. For example as shown in Ref.\cite{Refael}, the temporal
topological phenomena can be simulated by a qubit subjected to a two-frequency drive.
The Hamiltonian takes the form,
\begin{eqnarray}
H_{temp.}&=&v_1\cos (\omega _1t+\varphi _1)\sigma ^x+v_2\sin (\omega _2t+\varphi _2)\sigma ^y
\nonumber \\
&&+[m-b_1\cos (\omega _1t+\varphi _1 )-b_2\cos (\omega _2t
+\varphi _2)]\sigma ^z,
\nonumber \\
\end{eqnarray}
where the notations and the implication of this model can be found in Ref.\cite{Refael}.
This Hamiltonian corresponds to Eq.(\ref{controlH}), and can be realized by a
superconducting qubit. The rotation axis should be in arbitrary direction. The
topological phenomena are described by whether the whole Bloch sphere of the corresponding
states are covered or not, which is similar with our experiment performed.

\section*{Appendix E: Relation between dynamical quantum phase and dynamical Chern number}
In Ref. \cite{CG}, it is shown that the Loschmidt amplitude of a two-band system can be
written as
\begin{equation}\label{EqS1}
G(t)=\prod_k [\cos(|\mathbf{d}_f(k)|t)+i\hat{\mathbf{d}}_i(k)\cdot \hat{\mathbf{d}}_f(k)\sin(|\mathbf{d}_f(k)|t)],
\end{equation}
where $\mathbf{d}_i(k)$ and $\mathbf{d}_f(k)$ correspond to the pre- and post-quench Hamiltonians, respectively. The dynamical quantum phase transition (DQPT) occurs when the Loschmidt amplitude reaches zero at a critical time $t_c$. As we see from Eq. (\ref{EqS1}), the existence of zeroes of $G(t)$ requires that there are at least one critical momentum $k^{*}$ satisfying
\begin{equation}\label{EqS2}
\hat{\mathbf{d}}_i(k^{*})\cdot \hat{\mathbf{d}}_f(k^{*})=0,
\end{equation}
i.e., the vector $\hat{\mathbf{d}}_i(k)$ is perpendicular to $\hat{\mathbf{d}}_f(k)$ at the critical momentum $k^{*}$, and the DQPT occurs at
\begin{equation}\label{EqS3}
t_c=\frac{\pi}{|\mathbf{d}_f(k)|}(n+\frac{1}{2}),~~~~~~~~~~~n=0,1,2,\dots,
\end{equation}
and the Bloch vector satisfies $\hat{\mathbf{d}}(k^{*},t_c)=-\hat{\mathbf{d}}_i(k^{*})$ \cite{Chen}. For clarity, we employ the Ising model to elucidate the condition of DQPT. One has,
\begin{eqnarray}
{\bf d}_i(k)\cdot{\bf d}_f(k)&=&\sin ^2k+\cos ^2k-(g_i+g_f)\cos k+g_ig_f\nonumber\\
&=&0.
\end{eqnarray}
The solution exists when
\begin{eqnarray}
|\cos k|=\left| \frac{1+g_ig_f}{g_i+g_f}\right| < 1.
\end{eqnarray}
One can obtain $sgn[(1-|g_i|)(1-|g_f|)]=-1$, i.e. DQPT occurs if and only if the initial Hamiltonian and the final Hamiltonian belong to different phases for the Ising model, and $k^*=\pm \arccos \frac{1+g_ig_f}{g_i+g_f}$.

We also know from Refs. \cite{Chen} and \cite{Ueda} that a dynamical Chern number can be defined in momentum-time space in a quench process. First we should find the fixed points $k_m$ that satisfying $\hat{\mathbf{d}}_i(k_m)$ is parallel and anti-parallel to $\hat{\mathbf{d}}_f(k_m)$. Here we just focus on the transverse field Ising model, there are only two fixed points $k=0$ and $k=\pi$. Then the dynamical Chern number is defined as
\begin{equation}\label{EqS5}
C_{dyn}=\frac{1}{4\pi}\int_{0}^{\pi}dk\int_{0}^{\pi}dt^{\prime}(\hat{\mathbf{d}}\times\partial_{t^{\prime}} \hat{\mathbf{d}})\cdot\partial_k \hat{\mathbf{d}}
\end{equation}
where $t^{\prime}=\frac{t}{\mathbf{d}_f}$ is the rescaled time.

For the fixed point $k=0$, we have $\hat{\mathbf{d}}(0)=(0,sgn(g-1),0)$, and for the fixed point $k=\pi$, we have $\hat{\mathbf{d}}(\pi)=(0,sgn(g+1),0)$. The dynamical Chern number is calculated,
\begin{equation}\label{EqS6}
C_{dyn}=\frac{1}{2}(\cos\theta_{k=0}-\cos\theta_{k=\pi}),
\end{equation}
where $\theta_k$ is the induced angle between $\hat{\mathbf{d}}_i(k)$ and $\hat{\mathbf{d}}_f(k)$. In our experiment, we first choose $g_i=0.2$ and $g_f=1.5$, hence $\hat{\mathbf{d}}_i(0)\cdot\hat{\mathbf{d}}_f(0)=-1$ and $\hat{\mathbf{d}}_i(\pi)\cdot\hat{\mathbf{d}}_f(\pi)=1$, the dynamical Chern number is $C_{dyn}=-1$. As a result the Bloch sphere is fully covered as shown in Fig. 2(g) in the main text. From the continuity of the function $\hat{\mathbf{d}}_i(k)\cdot\hat{\mathbf{d}}_f(k)$, there must be a critical momentum $k^{*}$ between $0$ and $\pi$ satisfying $\hat{\mathbf{d}}_i(k^{*})\cdot\hat{\mathbf{d}}_f(k^{*})=0$, so we can draw a conclusion that the nontrivial dynamical Chern number ensures the occurrence of DQPT.

We also choose $g_i=0.2$ and $g_f=0.5$, we have $\hat{\mathbf{d}}_i(0)\cdot\hat{\mathbf{d}}_f(0)= \hat{\mathbf{d}}_i(\pi)\cdot\hat{\mathbf{d}}_f(\pi)=1$, and hence the dynamical Chern number $C_{dyn}=0$. In this case the Bloch sphere is not fully covered as shown in Fig. 2(n) in the main text, and the DQPT would not occur.

The nontrivial dynamical Chern number indicates the emergence of Skyrmion lattice in the momentum-time space. If $g_i=0.2$ and $g_f=1.5$, the dynamical Chern number is nontrivial, we consider the expectation value
\begin{equation}\label{EqS7}
\langle\hat{\mathbf{d}}(k,t)\rangle\equiv \langle\phi(k,t)|\hat{\mathbf{d}}\cdot\sigma|\phi(k,t)\rangle= -\hat{\mathbf{d}}(k,t)\cdot\hat{\mathbf{d}}_i(k).
\end{equation}
At $k=k^{*}$ and $t=t_c$, $\langle\hat{\mathbf{d}}(k,t)\rangle$ reaches the minimum $-1$ and $(k,t)=(k^{*},t_c)$ is the center in the texture of pseudospin as shown in Fig. 4(c). It forms a lattice during the time evolution with the lattice spacing is just the period of DQPT $\frac{\pi}{|\mathbf{d}_f(k)|}$. In the case $g_i=0.2$ and $g_f=0.5$, the DQPT would not occur, the dynamical Chern number is trivial and Skyrmion lattices would not appear as shown in Fig. 4(d) in the main text.

\section*{Appendix F: Error bar shown in the figures}

The dynamical free energy can be expressed in terms of $\hat{\textbf{d}}_i(k)$ and
	$\hat{\textbf{d}}(k,t)$
	\begin{equation}
		f(t) = -\frac{1}{N}\sum_k\log\frac{1+\hat{\textbf{d}}_i(k)\cdot\hat{\textbf{d}}(k,t)}{2},
	\end{equation}
	In our experimental setup, $\hat{\textbf{d}}_i(k)$ is a fixed unit vector. To estimate the experimental error
	of $f(t)$, we need only to estimate the fluctuation of $\hat{\textbf{d}}_i(k)$. Given a specific $k$,
	we have obtained $70$ state tomography data $\hat{\textbf{d}}(k,t)$ corresponding to different time points
	on the evolution path on the Bloch sphere. Each of these tomography data $\hat{\textbf{d}}(k,t)$ is an average
	of $5000$ raw data. We estimate the fluctuation of $\hat{\textbf{d}}(k,t)$ by estimating the fluctuation of
	the radius of the evolution path traced on the Bloch sphere. For each path, we choose three equally
	separated state points $\hat{\textbf{d}}(k,t)$ and calculate the radius determined. Thus for $k$, we
	obtain $22$ estimation of the evolution path. The magnitude $\left\|\Delta\hat{\textbf{d}}(k)\right\|$ of
	the fluctuation of $\hat{\textbf{d}}(k,t)$ is evaluated by the standard deviation of the $22$ estimation
	of the radius. The error of the dynamical free energy is hence
	\begin{equation}
		\Delta{f}(t)
		= \frac{1}{N}\sum_k\frac{\left\|\Delta\hat{\textbf{d}}(k)\right\|}{1+\hat{\textbf{d}}_i(k)\cdot\hat{\textbf{d}}(k,t)}.
	\end{equation}
Those error bars are indicated in the Fig. 3 and Fig. 4 in the main text.

\end{document}